\documentclass{rnaastex}

\begin{document}

\title{Comet C/2017~K2 (PANSTARRS): dynamically old or new?}

%% Note that the corresponding author command and emails has to come
%% before everything else. Also place all the emails in the \email
%% command instead of using multiple \email calls.
\correspondingauthor{Ra\'ul~de~la~Fuente~Marcos}
\email{rauldelafuentemarcos@ucm.es}

\author[0000-0002-5319-5716]{Ra\'ul~de~la~Fuente~Marcos}
\affiliation{AEGORA Research Group \\
             Facultad de Ciencias Matem\'aticas \\
             Universidad Complutense de Madrid \\
             Ciudad Universitaria, E-28040 Madrid, Spain}

\author[0000-0003-3894-8609]{Carlos~de~la~Fuente~Marcos}
\affiliation{Universidad Complutense de Madrid \\
             Ciudad Universitaria, E-28040 Madrid, Spain}

%% Note that RNAAS manuscripts DO NOT have abstracts.
\keywords{comets: general --- comets: individual (C/2017~K2)}

%% Start the main body of the article. If no sections in the 
%% research note leave the \section call blank to make the title.
\section{} 

At discovery time, C/2017~K2 (PANSTARRS) was the second most distant ($\approx$16~au) inbound active comet ever observed \citep{2017ApJ...849L...8M}. 
The fact that the comet was active beyond the crystallization zone was interpreted as evidence in favor of having an origin in the Oort 
Cloud and of being in the process of crossing the inner Solar System for the first time \citep{2017ApJ...847L..19J}. This conjecture was 
further explored by \citet{2018AJ....155...25H}, arriving to the conclusion that the comet was most probably from the Oort cloud (with a 
probability of having an interstellar provenance $\approx$1.7\%), but leaving open the answer to the question of the comet being dynamically 
new or old. \citet{2017ApJ...849L...8M} assumed that C/2017~K2 is a dynamically new Oort cloud comet, but \citet{2018arXiv180210380K} have 
carried out extensive calculations to conclude that it is dynamically old.

Here, we present numerical evidence suggesting that C/2017~K2 is dynamically old. Our results are based on the latest orbital solution
avaliable from \href{http://ssd.jpl.nasa.gov/sbdb.cgi}{JPL's Small-Body Database}, which is different from the ones used by 
\citet{2018AJ....155...25H} and \citet{2018arXiv180210380K}. Our full $N$-body calculations have been carried out as described by 
\citet{2012MNRAS.427..728D} ---with input state vectors computed as discussed by \citet{2015MNRAS.453.1288D}--- this approach and tools are
also different from the ones used by \citet{2018AJ....155...25H} and \citet{2018arXiv180210380K}. The current orbit determination of 
C/2017~K2 (epoch JDTDB~2457961.5, 27-July-2017) computed by D. Farnocchia (4-April-2018) is based on 491 observations for a data-arc span of 
1770~d and has perihelion distance, $q$=1.81066$\pm$0.00009~au, eccentricity, $e$=1.00033$\pm$0.00003, inclination, $i$=87\fdg55332$\pm$0\fdg00008, 
longitude of the ascending node, $\Omega$=88\fdg1855$\pm$0\fdg0007, and argument of perihelion, $\omega$=236\fdg022$\pm$0\fdg003; with an 
absolute magnitude of 6.3$\pm$0.8, C/2017~K2 may be under 18~km wide \citep{2017ApJ...847L..19J}.  

\begin{figure}[!ht]
\begin{center}
\includegraphics[scale=0.5,angle=0]{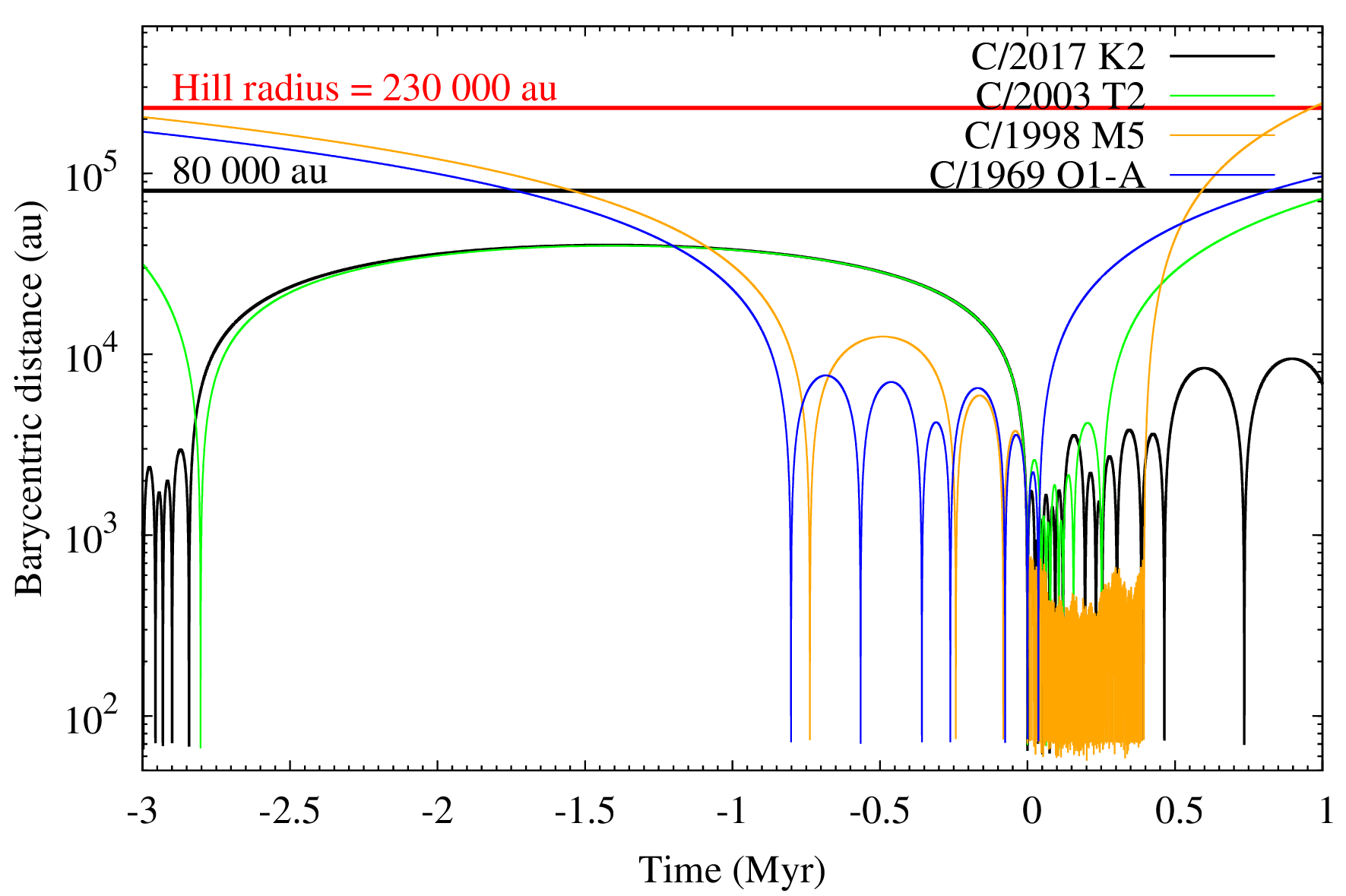}
\includegraphics[scale=0.5,angle=0]{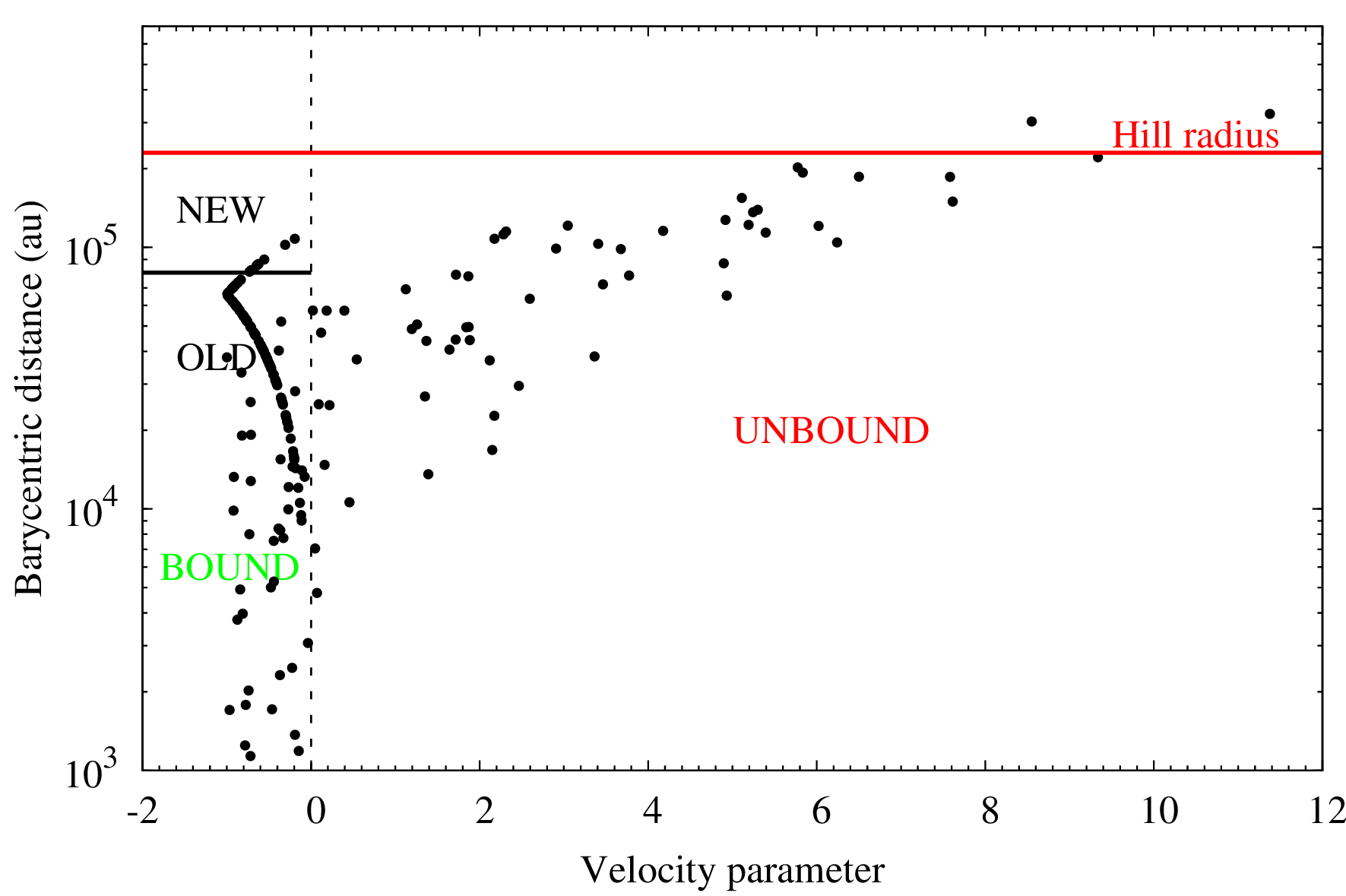}
\caption{Evolution of the barycentric distance of C/2017~K2, C/2003~T2, C/1998~M5, and C/1969~O1-A (nominal orbits, top panel); zero instant  
         of time, epoch JDTDB~2458200.5, 23-March-2018. Values of the barycentric distance as a function of the velocity parameter 3~Myr 
         into the past for 200 control orbits of C/2017~K2 (bottom panel). The velocity parameter is the difference between the barycentric 
         and escape velocities at the computed barycentric distance in units of the escape velocity. Positive values of the velocity 
         parameter identify control orbits that could be the result of capture from outside the Solar System. The thick black line 
         corresponds to the aphelion distance ---$a \ (1 + e)$, limiting case $e=1$--- that defines the domain of dynamically old comets 
         (i.e. dynamically old comets have $a^{-1} > 2.5 \times 10^{-5}$~au$^{-1}$, see \citealt{2017MNRAS.472.4634K}); the thick red line 
         signals the radius of the Hill sphere of the Solar System (see e.g. \citealt{1965SvA.....8..787C}). 
\label{fig:1}}
\end{center}
\end{figure}

The orbit determination of C/2017~K2 is hyperbolic at the 11$\sigma$ level, but it is somewhat similar in terms of $q$ and $i$ to those of 
the long-period comets C/1969~O1-A (Kohoutek), $q$=1.72~au, $e$=0.9991, $i$=86\fdg3, C/1998~M5 (LINEAR), $q$=1.74~au, $e$=0.9960, $i$=82\fdg2, 
and C/2003~T2 (LINEAR), $q$=1.79~au, $e$=0.9997, $i$=87\fdg5. Although the present-day orbital solution of C/2017~K2 is slightly hyperbolic, 
it may have followed an elliptical path in the past, and $N$-body simulations can confirm or reject this hypothesis. Figure~\ref{fig:1}, top 
panel, shows the evolution of the barycentric distance of the nominal orbit of the four comets; the output cadence is 100 yr. Consistent 
with the interpretation made by \citet{2018arXiv180210380K}, C/2017~K2 is not a dynamically new Oort cloud comet and its future evolution is 
rather chaotic; its past evolution matches that of C/2003~T2. The analysis of 200 control orbits of C/2017~K2 compatible with the 
observations and integrated backwards in time for 3~Myr shows (see Figure~\ref{fig:1}, bottom panel) that most of them, 67\%, are consistent 
with a bound and dynamically old comet, but about 29\% of the studied orbits are compatible with an interstellar origin. 

Our independent analysis confirms the results in \citet{2018arXiv180210380K}, but the probability of this object having an interstellar
origin seems to be significantly higher than that obtained by \citet{2018AJ....155...25H}, although their integrations only reach 1~Myr into
the past. Comet C/2017~K2 joins the growing sample of dynamically old Oort cloud comets that, in the case of currently hyperbolic ones, 
corresponds to the subsample of bound objects discussed by \citet{2018MNRAS.476L...1D}.

%% An example figure call using \includegraphics

%% An example table using AASTeX's deluxetable. Note that since
%% only one figure OR one table is allowed this is commented out.
%\begin{deluxetable}{ccl}
%\tablecaption{Example table some English and Greek letters\label{tab:1}}
%\tablehead{
%\colhead{Index number} & \colhead{English} & \colhead{Greek}
%}
%\startdata
%1 & a & alpha ($\alpha$) \\
%2 & b & beta ($\beta$) \\
%3 & c & gamma ($\gamma$) \\
%4 & d & delta ($\delta$) \\
%5 & e & epsilon ($\epsilon$) \\
%\enddata
%\tablecomments{Long tables should only show a short example with the full
%version as a machine readable table with the article.}
%\end{deluxetable}  

\acknowledgments

We thank S.~J. Aarseth for providing the code used in this research and A.~I. G\'omez de Castro for providing access to computing facilities. 
This work was partially supported by the Spanish MINECO under grant ESP2015-68908-R. In preparation of this Note, we made use of the NASA 
Astrophysics Data System and the MPC data server.

\end{document}